\documentclass[twocolumn,showpacs,preprintnumbers,amsmath,amssymb,superscriptaddress,prd]{revtex4}

\usepackage{graphicx}
\usepackage{dcolumn}
\usepackage{bm}
\usepackage{amsmath,amssymb}

\begin{document}

\preprint{APS/123-QED}

\title{Constraining Non-local Gravity by S2 star orbits}

\author{K. F. Dialektopoulos}
\affiliation{Dipartimento di Fisica "E. Pancini", Universit\`{a} di Napoli "Federico II", Compl. Univ. di Monte S. Angelo, Edificio G, Via Cinthia, I-80126, Napoli, Italy.} 
\affiliation{Istituto Nazionale di  Fisica Nucleare (INFN) Sez. di Napoli, Compl. Univ. di Monte S. Angelo, Edificio G, Via Cinthia, I-80126, Napoli, Italy.}

\author{D. Borka}
\email[Corresponding author:]{dusborka@vin.bg.ac.rs}
\affiliation{Atomic Physics Laboratory (040), Vin\v{c}a Institute of Nuclear Sciences, University of Belgrade, P.O. Box 522, 11001
Belgrade, Serbia.}

\author{S. Capozziello}
\affiliation{Dipartimento di Fisica "E. Pancini", Universit\`{a} di Napoli "Federico II", Compl. Univ. di Monte S. Angelo, Edificio G, Via Cinthia, I-80126, Napoli, Italy.} 
\affiliation{Istituto Nazionale di  Fisica Nucleare (INFN) Sez. di Napoli, Compl. Univ. di Monte S. Angelo, Edificio G, Via Cinthia, I-80126, Napoli, Italy.}
\affiliation{Tomsk State Pedagogical University, ul. Kievskaya, 60, 634061 Tomsk, Russia.}
\affiliation{Laboratory for Theoretical Cosmology,
Tomsk State University of Control Systems and Radioelectronics (TUSUR), 634050 Tomsk, Russia.	}

\author{V. Borka Jovanovi\'{c}}
\affiliation{Atomic Physics Laboratory (040), Vin\v{c}a Institute of Nuclear Sciences, University of Belgrade, P.O. Box 522, 11001
Belgrade, Serbia.}

\author{P. Jovanovi\'{c}}
\affiliation{Astronomical Observatory, Volgina 7, P.O. Box 74, 11060 
Belgrade, Serbia.}

\date{\today}

\begin{abstract}
Non-local theories of gravity have recently gained a lot of interest because they can suitably represent the behavior of gravitational interaction  in the ultraviolet regime. Furthermore, at  infrared scales,  they  give rise to notable cosmological effects which could be important to describe the dark energy behavior. In particular, exponential forms of the distortion function 
seem particularly useful for this purpose. 
Using Noether Symmetries, it can be shown that the only non-trivial 
form of the distortion function is the exponential one, which is working not only for  
cosmological mini-superspaces, but also in a spherically symmetric 
spacetime. Taking this result into account, we study the weak field 
approximation of this type of non-local gravity, and comparing with 
the orbits of S2 star around the Galactic center (NTT/VLT data), we 
set constraints on the parameters of the theory. Non-local effects do not play a significant role on the orbits of S2 
stars around Sgr A*, but give  richer phenomenology at 
cosmological scales than the $\Lambda$CDM model. Also, we show that 
non-local gravity model  gives better agreement between 
theory and astronomical observations than Keplerian orbits.
\end{abstract}


\maketitle

\section{Introduction}

It is well established  that General Relativity (GR), 
together with the associated \textit{concordance model} in 
cosmology, $\Lambda$CDM, are the most successful explanations for 
gravitational and cosmological effects in the Universe. They have both 
passed the observational tests with flying colors. Cosmic Microwave 
Background Radiation, supernovae type Ia, large scale structures, 
as well as Solar System experiments and galactic rotation curves are some 
of these tests. However, the inability to find a convincing 
explanation for the accelerated expansion of the Universe, the huge 
discrepancy between the theoretical and observed values of the 
cosmological constant at early and late times, the fact that no particle candidate for dark 
matter has been observed at fundamental scales, together with the failure to confirm the 
existence of supersymmetry at TeV scales, led the scientists to 
pursue  alternative explanations for the gravitational 
interaction. 

The list of modifications is huge; from adding new fields, e.g. 
scalar-tensor, galileons, Kinetic Gravity Braiding (KGB), 
quintessence, Tensor-vector-scalar gravity (TeVeS), massive gravity, 
bi-gravity and more, to higher-order theories, e.g. $f(R)$, 
$f(\mathcal{G})$, conformal gravity, to higher dimensional theories, 
e.g. Kaluza-Klein, Dvali-Gabadadze-Porrati (DGP), Randal-Sundrum, as 
well as to emergent approaches, such as Causal Dynamical 
Triangulation (CDT) or entropic gravity. For more details, the 
interested reader is refereed to the exhausting literature 
\cite{Curvature,capo11a,clif12,Nojiri:2010wj,Nojiri:2017ncd}. 

Among all the above, more than a decade ago, a non-local modification 
at infrared scales was proposed \cite{Deser:2007jk} to explain the 
late-time acceleration of the universe. Non-localities usually appear 
naturally in quantum loop corrections, as well as when one considers 
the effective action approach to sting/M-theory. It has also been 
proposed \cite{Donoghue:1994dn,Giddings:2006sj} that such terms could 
be considered as solution to the black hole information paradox.

During this decade many attempts have been done in the literature to 
study non-localities in various contexts 
\cite{univ4,modesto1,modesto2,st1,st2,loop,jm}. Bouncing solutions in 
the string theory framework are discussed in \cite{Arefeva:2007wvo}, 
while in \cite{Arefeva:2007xdy} they present phantom dark energy 
solutions to explain the accelerated expansion of the Universe. 
Non-Gaussianities during inflation are studied in 
\cite{Barnaby:2008fk}. Apart from the ultraviolet scales, a lot of 
progress has been done in the infrared scales too. Unification of 
inflation with late-time acceleration, as well as, the dynamics of a 
local form of the theory have been studied in 
\cite{Nojiri:2007uq,Jhingan:2008ym}. In \cite{Deser:2013uya}, they 
prove that non-local gravities are ghost-free and stable and that 
they do not alter the predictions of GR for gravitationally bound 
systems. Last but not least, in \cite{Deffayet:2009ca}, they try to 
fix the functional form of the distortion function, while in 
\cite{Koivisto:2008dh,Koivisto:2008xfa} they study the dynamics of 
the theory and its Newtonian limit. For a detailed review on the 
topic, we refer to \cite{Barvinsky:2014lja}.

In parallel, symmetries always played a significant role in field 
theories. It would be thus very desired, if not necessary, if any new 
proposed theory is invariant under specific transformations. It has 
been proposed \cite{Cimento,Gaetano, Sergey, Dialektopoulos:2018qoe,Tsamparlis:2018nyo}, that 
the Noether Symmetry Approach could be used as a selective criterion 
for gravitational models that are invariant under point 
transformations. It has been successfully studied in the literature 
numerous times 
\cite{Bahamonde:2017sdo,Capozziello:2016eaz,Capozziello:2018gms, 
Bahamonde:2018zcq,Bahamonde:2018ibz,Karpathopoulos:2017ebb, 
Paliathanasis:2017kzv,Dimakis:2017kwx,Paliathanasis:2014iva, 
Basilakos:2013rua}. It turns out that, apart from selecting theories 
of gravity, Noether symmetries of dynamical systems can help us 
calculate the invariant functions and use them to reduce the dynamics 
of the system and find analytical solutions.

In this paper, we consider the non-local theory proposed by Deser and 
Woodard but in its local representation. We apply the Noether 
Symmetry Approach in a spherically symmetric spacetime and find those 
functional forms of the \textit{distortion function}, that keep the 
point-like Lagrangian invariant. Similar analysis in the cosmological 
minisuperspace \cite{Bahamonde:2017sdo} has shown that the only 
possible forms are the linear and the exponential ones. The results 
included here are in complete agreement with those in cosmology. The 
linear form has been suggested \cite{Wetterich:1997bz} to cure the 
unboundedness of the Euclidean gravity action, while the exponential 
\cite{Nojiri:2017ncd} to explain the late-time acceleration, to unify 
the inflation era with the current one and more. However, up to now, 
they were both chosen by hand to explain phenomenology, while in 
\cite{Bahamonde:2017sdo} and also here, the form of the non-local modification  is chosen 
from first principles, that is  the existence of the Noether symmetry. 

Furthermore, we find the weak field limit of the theory with the 
exponential coupling and we also calculate the Post-Newtonian (PN) 
terms up to $g_{00}\sim \mathcal{O}(6)$. The local representation of 
this non-local model can be formulated as a biscalar-tensor theory. 
However, one of the two scalar fields is not dynamical. In the PN 
analysis, two new length scales arise, however, only one of them is 
physical; the other one belongs to the auxiliary degree of freedom 
introduced to localize the original action. 

Finally, we consider the orbits of S2 star around the Galactic 
center and, by comparing the PN terms of our theory with observations, 
we are able to set some bounds on the above dynamical length scale. 
S-stars are the bright stars which move around the centre of our
Galaxy \cite{ghez00,scho02,ghez08,gill09a,gill09b,genz10,gill12,meye12,gill17,hees17,chu17}
where the compact radio source Sagittarius A* (or Sgr A*) is 
located. For one of them, called S2, a deviation from its Keplerian 
orbit was observed \cite{gill09a,meye12,gill17,boeh17,hees17,chu17}, 
but the community debates to integrate its motion in the framework 
of GR.

Obviously, the non-localities are not expected to contribute 
significantly at astrophysical and galactic scales, because otherwise 
they would have been observed. However, what we see is that our 
approach is consistent with the orbits of S2 star around Sgr A* 
and thus we extend its range of validity, which up to now was only 
at cosmological scales, to the astrophysical ones too.

The present paper is organized as follows: in Sec. 
\ref{sec:non-local} we sketch the theory of non-local gravity and it 
biscalar-tensor representation. In Sec. \ref{sec:noether} we apply 
the Noether Symmetry Approach in a spherically symmetric spacetime 
and we find those theories that are invariant under point 
transformations. In Sec. \ref{sec:weakfield} we derive weak field 
limit of the exponential coupling, as well as Post-Newtonian 
corrections. In Sec. \ref{sec:simulations} we describe the 
simulations of stellar orbits in the gravitational potential and  the fitting procedure. An extended discussion about our 
results, together with future perspectives are presented in Sec. 
\ref{sec:discussion}. We draw conclusions in Sec. 
\ref{sec:conclusions}.

\section{Non-local Gravity}
\label{sec:non-local}

It has been more than a decade that Deser and Woodard 
\cite{Deser:2007jk} proposed a non-local modification of the 
Einstein-Hilbert action, which has the following form
\begin{equation}\label{actionDeser}
\mathcal{S} =\frac{1}{2 \kappa ^2} \int d^4 x \sqrt{-g} \left[ R 
\left( 1 + f(\square ^{-1} R) \right) \right]  \,,
\end{equation}
where $R$ is the Ricci scalar and $f(\square^{-1}R)$ is an arbitrary 
function, called \textit{distortion function}, of the non-local term 
$\square ^{-1}R$, which is explicitly given by the retarder Green's 
function
\begin{equation}\label{Greenfunc}
\mathcal{G}[f](x)=(\square ^{-1}f)(x) = \int d^4x' 
\sqrt{-g(x')}f(x')G(x,x')\,.
\end{equation}
Setting $f(\square^{-1}R)=0$, the above action is equivalent to 
the Einstein-Hilbert one.  The non-locality is introduced by the inverse 
of the d'Alembert operator.

A local representation of \eqref{actionDeser} has been proposed 
in \cite{Nojiri:2007uq}; they introduce two auxiliary scalar fields 
$\phi$ and $\xi$ and they rewrite the action \eqref{actionDeser} as
\begin{align}
\mathcal{S} &= \frac{1}{2 \kappa^2} \int d^4x \sqrt{-g} \left[R 
\left( 1+ f(\phi)\right) + \xi \left(\square \phi - R\right)\right]   \nonumber \\
&= \frac{1}{2 \kappa^2} \int d^4x \sqrt{-g} \left[R \left( 1+ 
f(\phi) -\xi\right) - \nabla^{\alpha}\xi \nabla_{\alpha} \phi  
\right] \,, \label{actionscalar}
\end{align}
where we just integrated out a total derivative. By varying the 
action with respect to $\xi$ and $\phi$ respectively, we get
\begin{align}\label{eqphi}
\square \phi &= R  \Rightarrow \phi = \square ^{-1} R \,,\\
\square \xi &= -R \frac{df}{d\phi}\,, \label{eqxi}
\end{align}
where the equation \eqref{eqphi} is just a constraint to recover 
\eqref{actionDeser}, but the equation \eqref{eqxi} is a non-trivial 
dynamical equation for $\xi$. Moreover, variation of the action 
\eqref{actionscalar} with respect to the metric yields,
\begin{align}
\Big( 1 &+ f(\phi) - \xi \Big) G_{\mu\nu}+ \frac{1}{2}g_{\mu\nu} \nabla^{\alpha} \xi \nabla_{\alpha} \phi  =  \nonumber \\ 
\label{metriceq}
= &\kappa^2 T_{\mu\nu} ^{M} + \nabla_{\mu} \xi \nabla_{\nu} \phi + \left( \nabla_{\mu}\nabla_{\nu} - g_{\mu\nu} \square\right) \left( f(\phi) - \xi \right)\,.
\end{align} 
Another interesting equation is the trace of \eqref{metriceq} which, 
after the use of \eqref{eqphi},\eqref{eqxi}, reads
\begin{align}\label{R}
\left( 1 + f(\phi) - \xi - 6 f'(\phi)\right) &R = \nonumber \\
=-\kappa^2 T ^{M} + \nabla_{\alpha} \xi \nabla^{\alpha} \phi &+ 3 
f''(\phi) \nabla_{\alpha} \phi \nabla^{\alpha} \phi  \,.
\end{align} 
In the next section, we will use the Noether Symmetry Approach to 
select the form of the theory, i.e. the distortion function, in order 
for it to be invariant under point transformations. As we will see 
only the linear and the exponential forms will survive; the only ones 
that were interesting in the literature up to now.

\section{Noether Symmetries in Non-local gravity}
\label{sec:noether}

Noether symmetries of second order differential equations can be 
connected to 
the collinations of the underlying manifold where the motion occurs. 
Thus, they can be used as a geometric criterion to determine the 
symmetries of dynamical systems, find the associated invariant 
functions and use them to reduce the dynamics of the system in order 
to find exact solutions. 

The {\it Noether Symmetry Approach} \cite{Cimento} has been extensively used in the 
literature to study the symmetries of several modified theories of 
gravity. The method goes as follows: we select a symmetry for the 
background spacetime which, in our case, is spherically symmetric. 
The metric is given by the following line element  
\begin{equation}\label{trmetric}
ds^2 = e^{\nu(t,r)}dt^2 - e^{\lambda(t,r)}dr^2 - r^2 d\Omega ^2\,,
\end{equation}
where $\nu(t,r)$ and $\lambda(t,r)$ are two arbitrary function 
which depend both on time $t$ and the radial coordinate $r$, since we 
do not know \textit{a priori} if  Birkhoff's theorem holds in 
non-local gravity.

Then, we substitute the metric \eqref{trmetric} into the Lagrangian 
density \eqref{actionscalar} and after integrating out all the total 
derivative terms, we obtain the point-like Lagrangian which, here, 
reads
\begin{align}
\mathcal{L} = &e^{-\frac{1}{2} (\lambda +\nu )} \Big(-e^{\nu } r^2 \nu _r \phi _r f'(\phi )+e^{\lambda } r^2 \lambda _t \phi _t f'(\phi )-\nonumber \\
&-2 e^{\nu } f(\phi ) \left(e^{\lambda }+r \lambda _r-1\right)-2 e^{\lambda +\nu }+2 e^{\nu }+e^{\nu } r^2 \xi _r \phi _r+\nonumber \\
&+e^{\nu } r^2 \nu _r \xi _r-e^{\lambda } r^2 \xi _t \phi _t-e^{\lambda } r^2 \lambda _t \xi _t+\nonumber \\
&+2 e^{\nu } \xi  \left(e^{\lambda }+r \lambda _r-1\right)-2 e^{\nu } r \lambda _r\Big)\,,\label{pointlag}
\end{align}
where the subscript denotes differentiation with respect to the variable.

The Noether vector, or else the generator of the point transformations, takes the form
\begin{align}
X = &\xi ^t(t,r,\nu,\lambda,\phi,\xi) \partial _t + \xi ^r(t,r,\nu,\lambda,\phi,\xi) \partial _r + \nonumber \\
&+ \eta ^{\nu}(t,r,\nu,\lambda,\phi,\xi) \partial _{\nu} +\eta ^{\lambda}(t,r,\nu,\lambda,\phi,\xi) \partial _{\lambda} +\nonumber \\
&+ \eta ^{\phi}(t,r,\nu,\lambda,\phi,\xi) \partial _{\phi} +\eta  ^{\xi}(t,r,\nu,\lambda,\phi,\xi) \partial _{\xi}\,. 
\end{align}
and in order for the dynamical system described by \eqref{pointlag} to 
have symmetries the following condition \cite{Dialektopoulos:2018qoe} has to be satisfied
\begin{equation}\label{noethercond}
X^{[1]}\mathcal{L} + \mathcal{L}\left( \frac{d\xi ^t}{dt} + \frac{d\xi^r}{dr}\right)  = \frac{dh^t}{dt}+\frac{dh^r}{dr}\,,
\end{equation}
where $h^t$ and $h^r$ are two arbitrary functions depending on $(t,r,\nu,\lambda,\phi,\xi)$. Expanding the above condition, we find a system of 75 equations with 9 unknown variables, i.e. 6  coefficients of the Noether vector $\{\xi^t,\,\xi^r,\,\eta^{\nu},\,\eta^{\lambda},\,\eta^{\phi},\,\eta^{\xi}\}$, 2 unknown functions in the right hand side of \eqref{noethercond}, $\{h^t,\,h^r\}$ and the form of the distortion function $f(\phi)$. Solving the system we find two possible models that are invariant under point transformations, that is 
\begin{equation}\label{f}
f(\phi ) = c_4 + c_3 \phi, \,\,\, \text{and}\,\,\, f(\phi) = c_4 + \frac{c_5}{c_1}
 e^{c_1 \phi}\,.
\end{equation}
Their symmetries are given by the following vectors respectively
\begin{align}
X = \left(c_1 t+\xi^t(r)\right)\partial _t -2 c_1 \partial _{\nu} &+\left(c_2+2c_1\right)\partial _{\phi}+ \nonumber \\&
+\left(c_3(c_2+2c_1)\right)\partial _{\xi} \,, \label{NV1}
\end{align}
\begin{align}
X = \left(c_2 t +\xi^t(r)\right)\partial _t -\frac{c_3}{2}r&\partial _r -\left(2c_2+c_3\right)\partial _{\nu} +c_1 c_3\partial _{\phi} +\nonumber \\
&+\left(c_3 (\xi-c_4-1)\right)\partial _{\xi}, \label{NV2} 
\end{align}
and in both cases, the functions in the right hand side of 
\eqref{noethercond} 
are arbitrary functions of $(t,r)$. The associated invariant function 
of each symmetry is given by
\begin{equation}
I= \left( \xi^t + \xi^r\right) \left(\dot{q}^i\frac{\partial \mathcal{L}}{\partial \dot{q}^i} -\mathcal{L}\right) - \eta ^i \frac{\partial \mathcal{L}}{\partial q^i} + h^t+h^r\,, 
\end{equation}
where $q$ are the variables of the configuration space, which, in our case,  is $\mathcal{Q} = \{ \nu,\lambda,\phi,\xi\}$. 

For the sake of completeness we have to  say that, from the Noether vectors 
\eqref{NV1} and \eqref{NV2}, 
one can construct the following Lagrange system
\begin{equation}
\frac{dt}{\xi^t} = \frac{dr}{\xi^r} = \frac{d\nu}{\eta^{\nu}} = \frac{d\lambda}{\eta^{\lambda}} = \frac{d\phi}{\eta^{\phi}} = \frac{d\xi}{\eta^{\xi}}\,,
\end{equation}
solve for each variable and find the so-called $0^{\text{th}}$ order 
invariants. 
Substituting these in the Euler-Lagrange equations given by 
\eqref{pointlag},  one can reduce the dynamics of the system and find 
exact spherically symmetric solutions. However, the point of this 
paper is to use the above forms of the distortion function and to 
study its weak field limit. This  is what we are going to do in the following  section.

\section{Weak field approximation}
\label{sec:weakfield}

We consider the exponential form for the distortion function, given 
by 
\eqref{f},  and we derive the non-local gravity potential in the weak 
field limit to test the orbit of the S2 star against it. Then, we compare 
the results with the set of S2 star orbit observations obtained by 
New Technology Telescope/Very Large Telescope (NTT/VLT). This study 
is a continuation of our previous studies where we considered various 
gravity models 
\cite{bork12,bork13,capo14,zakh14,bork16,bork16a,capo17,zakh16,zakh18}. 

It is well known from GR that, in order to recover 
the Newtonian potential for time-like particles \footnote{Here we 
refer only in the cases where, the matter-fields are only minimally 
coupled to the metric and to no other fields. This is also the case 
for the non-local theory under study.} we have to expand the $g_{00}$ 
component of the metric to $\Phi \sim \upsilon ^2  \sim 
\mathcal{O}(2)$, where $\Phi$ is the Newtonian potential and 
$\upsilon $ is the 3-velocity of a fluid element. If we want to 
study the PN limit we have to expand the components of 
the metric as
\begin{equation}
g_{00}\sim \mathcal{O}(6) \,,\,\,g_{0i} \sim \mathcal{O}(5) 
\,\,\text{and}\,\, g_{ij} \sim \mathcal{O}(4)\,.
\end{equation}
Obviously, for the lowest order of the PN approximation 
we do not have to go up to $\mathcal{O}(6)$. However, as we would 
expect, two new length scales arise, which are related to the scalar 
degrees of freedom and thus we have to compute higher order 
corrections.

We want to study the behavior of the gravitational field generated 
by a point-like source and we consider that the metric is static and 
spherically symmetric. Before  proceeding, it is worth to make the 
following comment; even though in principle, we do not expect that 
 Birkhoff's theorem is valid in  non-local gravity, and that is 
the reason why, in order to derive the Noether symmetries, we 
considered a time-dependent line element, it is reasonable to believe 
that, as a first approximation in weak-field gravity, a static and 
spherically symmetric metric works as well. With this position, the metric assumes
the form
\begin{equation}\label{ssmetric}
ds^2 = A(r) dt^2 - B(r) dr^2 - r^2 d\Omega^2 \,.
\end{equation}
Although, we could take as fact that $B(r) = 1/A(r)$, in alternative 
theories of gravity, this cannot be chosen \textit{a priori}, since 
the existence of such solutions is not necessary. 

Obviously, since the metric \eqref{ssmetric} depends only on the 
radial coordinate, the scalar fields inherit the isometries of the 
metric and thus we have $\phi = \phi (r)$ and $\xi = \xi(r)$. The 
expansion of the metric components, as well as the scalar fields, 
read
\begin{subequations}
\begin{align}\label{Apert}
A(r) &= 1 + \frac{1}{c^2}\Phi(r) ^{(2)} + \frac{1}{c^4}\Phi (r) 
^{(4)} + \frac{1}{c^6}\Phi(r)^{(6)}+\mathcal{O}(8)\,,\\ \label{Bpert}
B(r) &= 1 + \frac{1}{c^2}\Psi(r) ^{(2)}  + 
\frac{1}{c^4}\Psi(r)^{(4)} + \mathcal{O}(6)\,,\\ \label{phipert}
\phi (r) &= \phi_0 + \frac{1}{c^2}\phi(r) ^{(2)} + \frac{1}{c^4}\phi 
(r) ^{(4)}  + \frac{1}{c^6}\phi(r)^{(6)}+\mathcal{O}(8) \,,\\ 
\label{xipert}
\xi (r) &= \xi_0 + \frac{1}{c^2}\xi(r) ^{(2)} + \frac{1}{c^4}\xi (r) 
^{(4)} + \frac{1}{c^6}\xi(r)^{(6)}+\mathcal{O}(8) \,,
\end{align}
\end{subequations}
where $\phi_0$ and $\xi_0$ are the constant background values of 
each 
field \footnote{It is easy to check that these constant scalar 
fields together with the Minkowski background metric, consist a 
solution of the equations \eqref{eqphi}-\eqref{metriceq}.}.

If we substitute the exponential form \eqref{f} for 
$f(\phi)$, i.e. $f = 1+e^{\phi}$ \footnote{Since they are arbitrary, we choose $c_1=1=c_4 = c_5$, to simplify the model. In addition, in order to recover the usual coupling of the Newton's constant with the Ricci scalar, we choose $\xi_0 = e^{\phi_0}$ and also $\phi_0 = 0$.}, we get the following four equations: the $00-$ and $11-$ components of \eqref{metriceq} and the two equations of the two scalar fields, \eqref{eqphi} and \eqref{eqxi} respectively

\begin{widetext}
\begin{align}\label{eq1}
2 B^2 \left(-\xi +e^{\phi }+2\right)+r B' \left(-2 \xi -r \xi '+r e^{\phi } \phi '+2 e^{\phi }+4\right)-\nonumber \\
-B \left(-2 \xi +2 \left(-r^2 \xi ''+r^2 e^{\phi } \phi ''+r^2 e^{\phi } \left(\phi '\right)^2+2 r e^{\phi } \phi '+e^{\phi }+2\right)+r \xi ' \left(r \phi '-4\right)\right)=0\,,\\ \label{eq2}
r A' \left(-2 \xi -r \xi '+r e^{\phi } \phi '+2 e^{\phi }+4\right)-\nonumber \\
-A \left(2 B \left(-\xi +e^{\phi }+2\right)+2 \xi +r^2 \xi ' \phi '+4 r \xi '-4 r e^{\phi } \phi '-2 e^{\phi }-4\right) = 0\,,\\ \label{eq3}
A^2 \left(-4 B^2 e^{\phi }+r B' \left(r \xi '-4 e^{\phi }\right)+B \left(-2 r^2 \xi ''-4 r \xi '+4 e^{\phi }\right)\right)+B r^2 \left(-e^{\phi }\right) \left(A'\right)^2+\nonumber \\
+A r \left(B \left(2 r e^{\phi } A''+A' \left(4 e^{\phi }-r \xi '\right)\right)-r e^{\phi } A' B'\right) = 0\,,\\ \label{eq4}
A^2 \left(-4 B^2-r B' \left(r \phi '+4\right)+2 B \left(r^2 \phi ''+2 r \phi '+2\right)\right)+B \left(-r^2\right) \left(A'\right)^2+\nonumber \\
+A r \left(B \left(2 r A''+A' \left(r \phi '+4\right)\right)-r A' B'\right) = 0\,.
\end{align}
\end{widetext}

Plugging the perturbations \eqref{Apert}-\eqref{xipert} into the above Eqs. \eqref{eq1}-\eqref{eq4}, we obtain three systems of four 
equations, one for each order, $\mathcal{O}(2),\,\mathcal{O}(4)$ and $\mathcal{O}(6)$. Since $B$ is calculated up to order $\mathcal{O}(4)$, in the last system, one of the equations will be a constraint to fix arbitrary integration constants. The solutions have the form
\begin{widetext}
\begin{subequations}
\begin{gather}\label{A(r)}
A(r) = 1 - \frac{2 G_N M \phi _c}{c^2 r} +\frac{G_N^2 M^2}{c^4 
r^2}\left[\frac{14 }{9}\phi _c^2 + \frac{18 r_{\xi }-11 r_{\phi }}{6 
r_{\xi } r_{\phi }} r \right] - \frac{G_N^3 M^3}{c^6 r^3} 
\left[\frac{50 r_{\xi }-7 r_{\phi }}{12 r_{\xi } r_{\phi }} \phi _c 
r+\frac{16 \phi _c^3}{27}-\frac{r^2 \left(2 r_{\xi }^2-r_{\phi 
}^2\right)}{r_{\xi }^2 r_{\phi }^2}\right] \,,\\ \label{B(r)}
B(r) = 1 + \frac{2 G_N M \phi_c}{3 c^2 r}+ \frac{G_N^2 M^2 }{c^4 
r^2}\left[\frac{2 \phi _c^2}{9} + \left(\frac{3}{2 r_{\xi 
}}-\frac{1}{r_{\phi }}\right)r\right]  \,,\\ \label{phi(r)}
\phi(r) =  \frac{4 G_N M \phi _c}{3 c^2 r} - \frac{G_N ^2 M^2}{c^4 
r^2} \left[\left(\frac{11}{6 r_{\xi }}+\frac{1}{r_{\phi }}\right) r - 
\frac{2 \phi _c^2}{9}\right] - \frac{G_N^3 M^3}{c^6 r^3}\left[ 
\frac{r^2}{r_{\phi }^2}  - \left( \frac{25}{12 r_{\xi }}-\frac{7}{6 
r_{\phi }}\right) \phi_c r-\frac{4 \phi _c^3}{81} \right]\,,\\ 
\label{xi(r)}
\xi(r) = 1 +\frac{G_N^2 M^2}{c^4 r^2} \left[\frac{2 \phi _c ^2}{3}- 
\left( \frac{13}{6 r_{\xi }}-\frac{1}{r_{\phi }}\right) r\right] + 
\frac{G_N^3 M^3}{c^6 r^3} \left[ \frac{20 \phi _c ^3}{27} - 
\left(\frac{1}{r_{\xi }^2}-\frac{1}{r_{\phi }^2}\right)r^2  
-\left(\frac{131}{36 r_{\xi }}+\frac{1}{6 r_{\phi }}\right)\phi _c r 
\right]\,.
\end{gather}
\end{subequations}
\end{widetext}
\noindent Here $\phi_c$ is a dimensionless constant and thus the effective 
gravitational coupling is $G_{\text{eff}} = G_N \phi _c$. Moreover, 
we see that two new length scales arise in the $\mathcal{O}(4)$ 
order. These are related to the two scalar degrees of freedom, $\phi$ 
and $\xi$ and thus to the non-localities. They are denoted as 
$r_{\phi}$ and $r_{\xi}$ respectively.

\section{Simulated orbits of S2 star in Non-local gravity potential}
\label{sec:simulations}

In order to constrain the free parameters, $\phi_c,\, r_{\phi}$ and 
$r_{\xi}$ we have to consider  the orbit of S2 star around the 
Galactic centre and fit them to astronomic observations by NTT/VLT. 
To do this we will need from the previous results the gravitational 
potential of the $g_{00}$ component of the metric, i.e. $A(r)$, 
\eqref{A(r)}. Following the expansion \eqref{Apert}, we identify
\begin{align}
\Phi^{(2)} (r) &= - \frac{2 G_N M}{r}\phi_c \,,\\
\Phi^{(4)} (r) &= \frac{G_N^2 M^2}{r^2} \left[\frac{14 }{9}\phi _c^2 
+ \frac{18 r_{\xi }-11 r_{\phi }}{6 r_{\xi } r_{\phi }} r 
\right]\,,\\
\Phi^{(6)} (r) &=  \frac{G_N ^3 M^3}{r^3} \left[\frac{7 r_{\phi }-50 
r_{\xi }}{12 r_{\xi } r_{\phi }} \phi _c r-\frac{16 \phi 
_c^3}{27}+\frac{2 r_{\xi }^2-r_{\phi }^2}{r_{\xi }^2 r_{\phi 
}^2}r^2\right]
\end{align}
We want to determine the free parameters of the theory, $\phi_c, \, 
r_{\phi}$ and $r_{\xi}$. We take specific values for 
$\phi_c$ = 1 (in order to obtain the Newtonian limit), and fix the 
parameter space of the other two.

Our aim is to determine these parameters using astrometric observations of 
S2 star orbit. In order to constrain parameters $r_{\phi}$ and $r_{\xi}$ 
by astronomical observations, we performed two-body 
simulations in non-local gravity potential

\begin{equation}
\label{eqnmoto}
\mathbf{\dot{r}}=\mathbf{v},\hspace*{0.5cm}
\mu\mathbf{\ddot{r}}=-\triangledown{U_{NL}}\left(
\mathbf{r}\right),
\end{equation}

\noindent where $\mu={M\cdot m_S}/({M+m_S})$ is the reduced mass in
the two-body problem.

The positions of the S2 star along its true orbit are calculated at
the observed epochs using two-body simulations in the non-local 
gravity potential, assuming that distance to the S2 star is $d$ = 8.3 
kpc and mass of central black hole $M_{BH}$=4.3 $\times 10^6 M_\odot$
\cite{gill09b}. In order to compare them with observed
positions we have to calculate the corresponding apparent orbits 
($x$, $y$) \cite{bork13}. The mass $M_{BH}$ of central object can be obtained independently using different observational techniques, such as e.g. virial analysis of the ionized gas in the central parsec \cite{lacy82}, $M-\sigma$ (mass - bulge velocity dispersion), the relationship for the Milky Way \cite{trem02}, or from orbits of S-stars \cite{gill09a,gill09b}. In the latter case, the mass of the SMBH was estimated using 2-body and N-body Keplerian and general relativistic orbit models (see \cite{gill17}). Inspite the fact that relativistic 2-body models resulted with slightly bigger values for both $M_{BH}$ and $d$, it was not possible to obtain the stastistically significant difference between these estimates, nor to detect any of the leading-order relativistic effects \cite{gill17}. Similarly, it would be also the case with mass estimates obtained by our 2-body simulations in non-local gravity. Therefore, in our simulations we used the statistically most significant estimates obtained from combined Keplerian orbit fit of 17 S-stars, which were also in agreement with a corresponding results determined from the statistical cluster parallax (see \cite{gill17}). Since our goal was not to make a new estimate of mass $M_{BH}$ using non-local gravity, but instead studying the possible deviations from Keplerian orbit of S2 star (which could indicate signatures for non-local gravity on these scales), we adopted the above estimates for the mass of central object ($M_{BH} = 4.3 \times 10^6 M_\odot$), as well as the distance to the S2 star given by \cite{gill09a,gill09b} ($d$ = 8.3 kpc), and constrained only the remaining two free parameters of non-local gravity potential ($r_\phi$, $r_\xi$). One should also note that slightly different masses would effect the values of precession angle but not significantly.

We vary the parameters $r_{\phi}$ and $r_{\xi}$ over some intervals, 
and search for those solutions which for the simulated orbits in non-local
gravity give at least the same ($\chi^{2}$ = 1.89) or better fits
($\chi^{2} < 1.89$) than the Keplerian orbits. 

We are simulating orbit of S2 star in the non-local 
gravity potential by numerical integration of equations of motion. We 
perform fitting using LMDIF1 routine from MINPACK-1 Fortran 77 
library which solves the nonlinear least squares problems by a 
modification of Marquardt-Levenberg algorithm \cite{more80,bork13}, 
according to the following procedure:

\begin{enumerate}
\item We start the first iteration using a guess of initial 
position $(x_0, y_0)$ and velocity $(\dot{x}_0, \dot{y}_0)$ of S2 
star in the orbital plane (true orbit) at the epoch of the first 
observation;
\item the true positions $(x_i, y_i)$ and velocities $(\dot{x}_i,
\dot{y}_i)$ at all successive observed epochs are then calculated by
numerical integration of equations of motion, and projected into
the corresponding positions $(x_i^c, y_i^c)$ in the observed plane
(apparent orbit);
\item in order to obtain discrepancy between the simulated and 
observed apparent orbit, we estimate the reduced $\chi^{2}$:
\begin{equation}
    \chi^{2} =
    \dfrac{1}{2N-\nu}{\sum\limits_{i = 1}^N {\left[ {{{\left(
    {\dfrac{x_i^o - x_i^c}{\sigma_{xi}}} \right)}^2}
    + {{\left( \dfrac{y_i^o - y_i^c}{\sigma_{yi}}
    \right)}^2}} \right]} },
\label{chi2}
\end{equation}
where $(x_i^o, y_i^o)$ and $(x_i^c, y_i^c)$ are the corresponding observed and
calculated apparent positions, $N$ is the number of observations, $\nu$ is
number of initial conditions (in our case $\nu=4$), $\sigma_{xi}$ and
$\sigma_{yi}$ are uncertainties of observed positions;
\item the new initial conditions are estimated by the fitting routine and the
steps 2-3 are repeated until the fit is converging, i.e. until the minimum of
reduced $\chi^{2}$ is achieved.
\end{enumerate}

For more detailed description about fitting procedure and numerics see in papers 
\cite{more80,bork13}.

\section{Results and Discussion}
\label{sec:discussion}

\begin{figure}[ht!]
\centering
\includegraphics[width=0.49\textwidth]{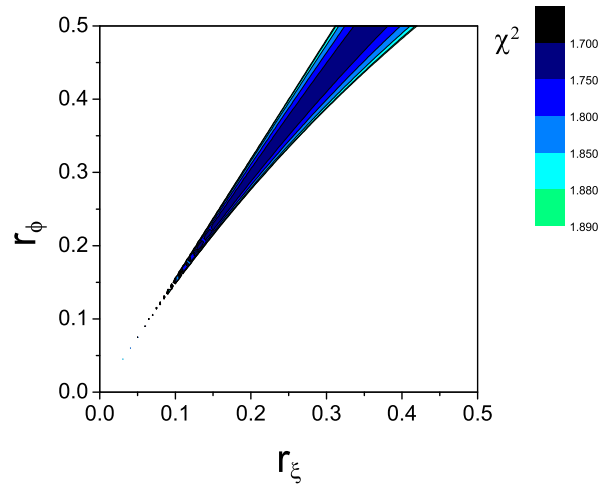}
\caption{(Color online) The maps of the reduced $\chi^{2}$ over the
$r_{\phi}-r_{\xi}$ parameter space (in $AU$) for all simulated orbits of 
S2 star which give at least the same or better fits than the 
Keplerian orbits ($\chi^{2}$ = 1.89). With a decreasing value of $\chi^{2}$ 
(better fit) colors in grey scale are darker. A few contours are presented for 
specific values of reduced $\chi^{2}$ given in the figure's legend.}
\label{fig01}
\end{figure}

\begin{figure*}[ht!]
\centering
\includegraphics[width=0.49\textwidth]{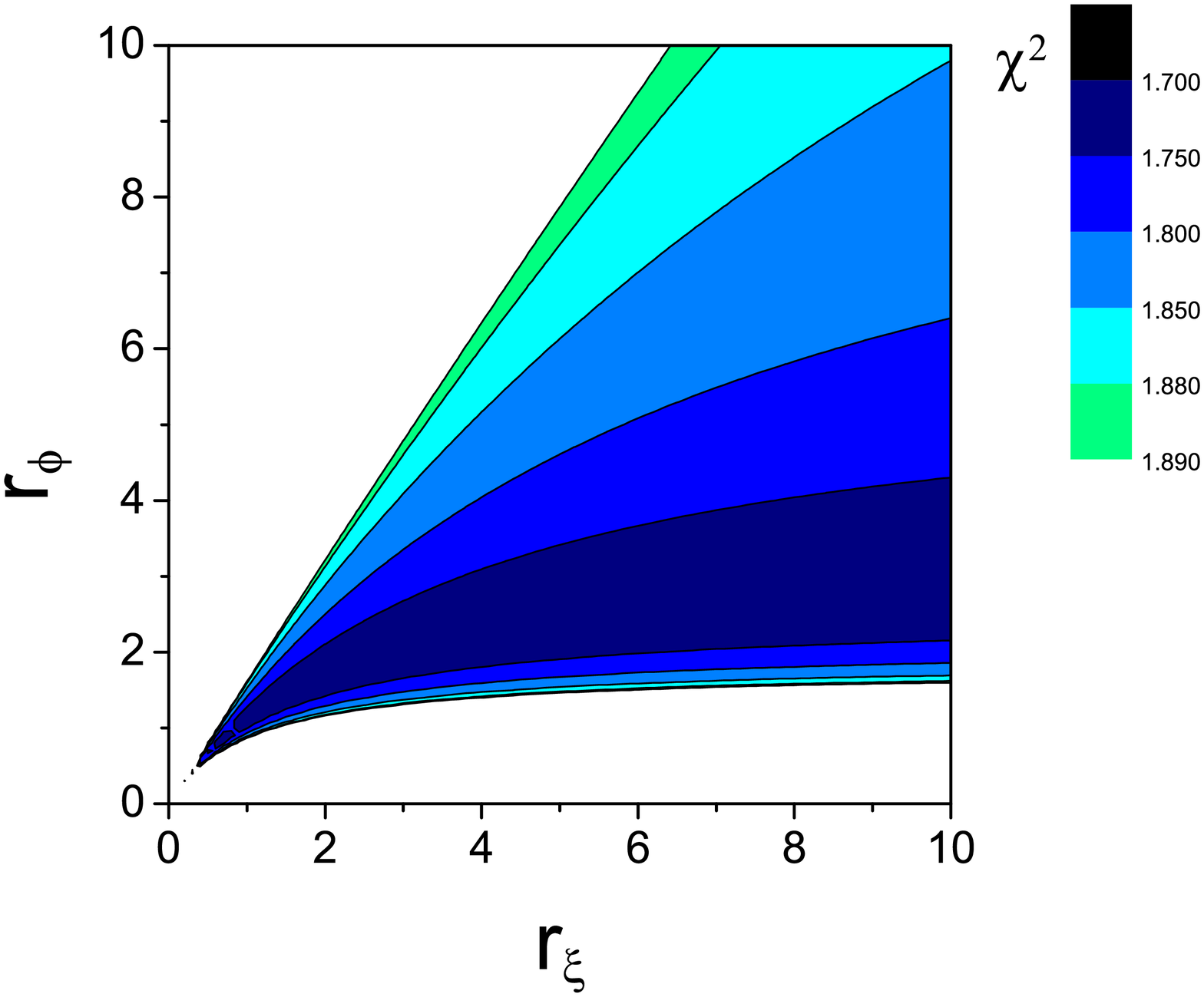}
\includegraphics[width=0.49\textwidth]{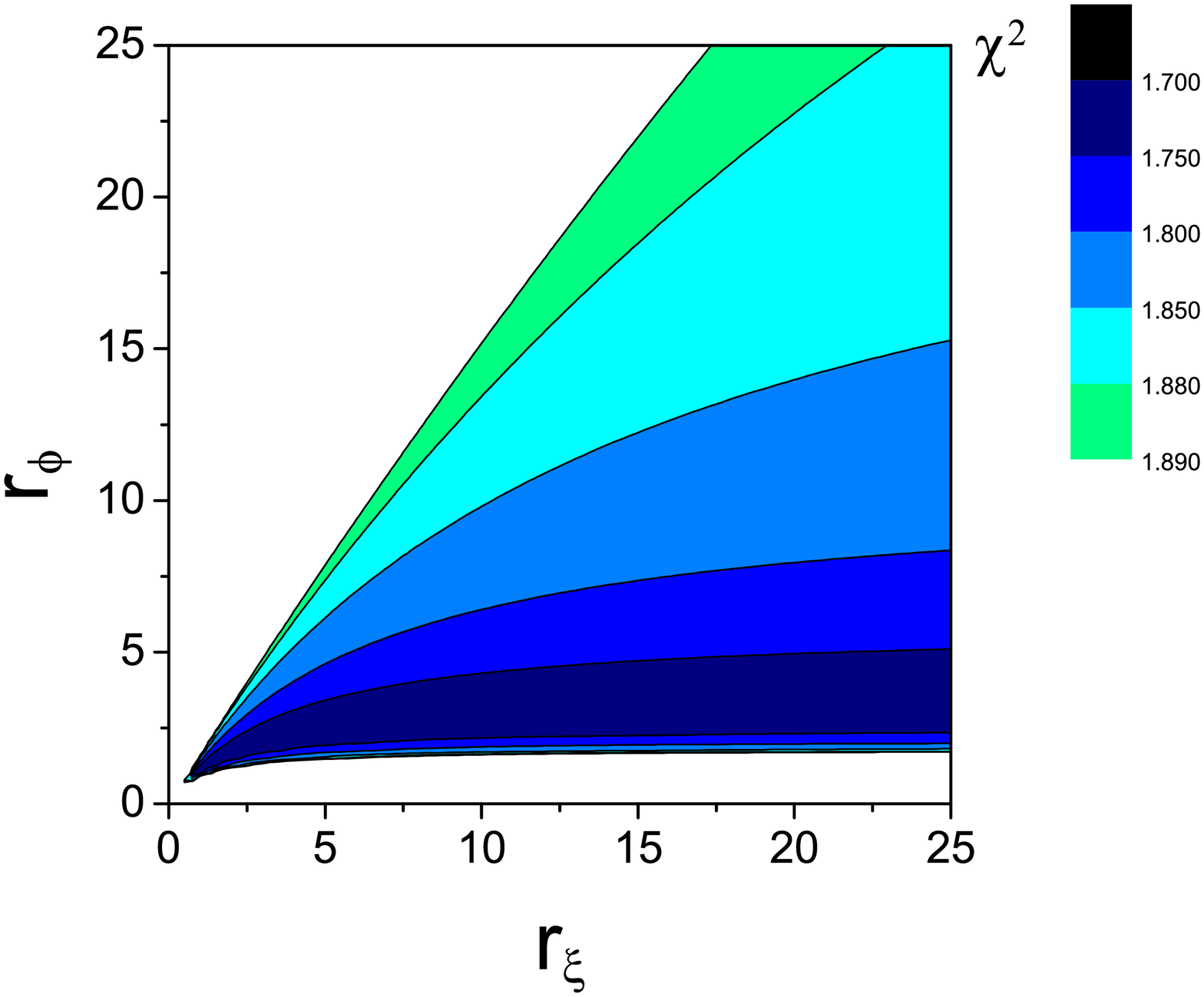}
\includegraphics[width=0.49\textwidth]{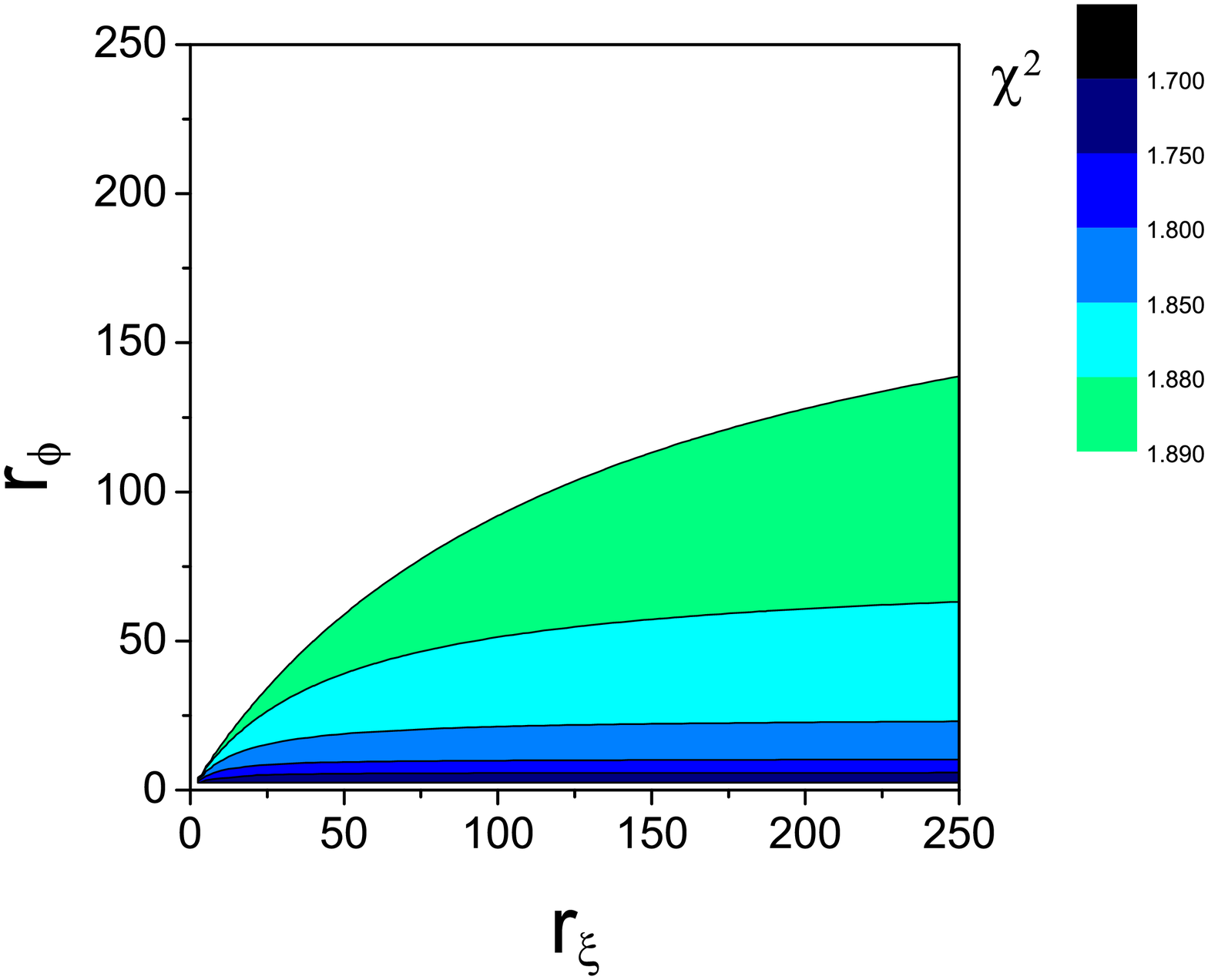}
\includegraphics[width=0.49\textwidth]{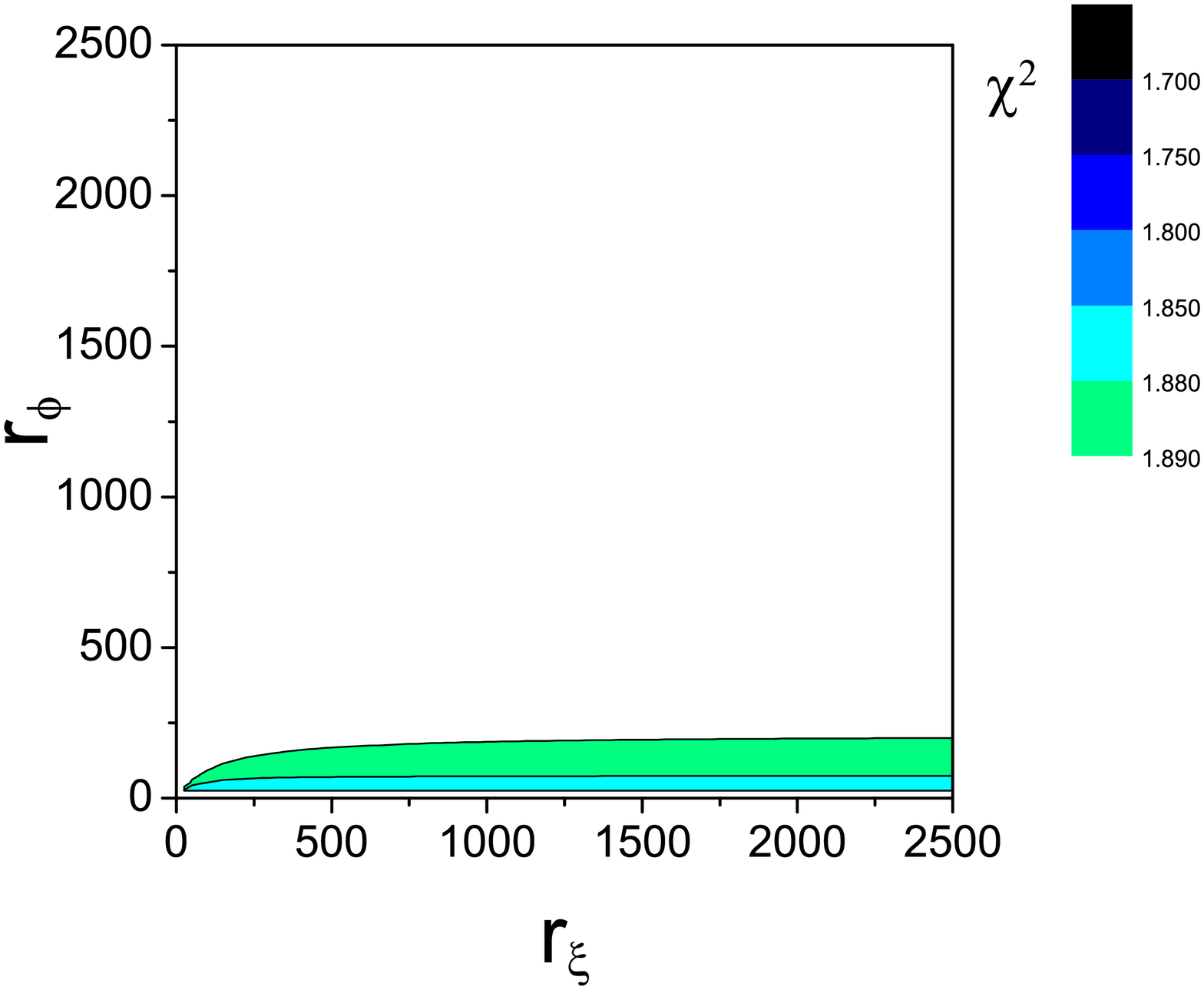}
\caption{The same like figure 1, but for more extended region of 
$r_{\phi}-r_{\xi}$ parameter space.}
\label{fig02}
\end{figure*}

\begin{figure*}[ht!]
\centering
\includegraphics[width=0.40\textwidth]{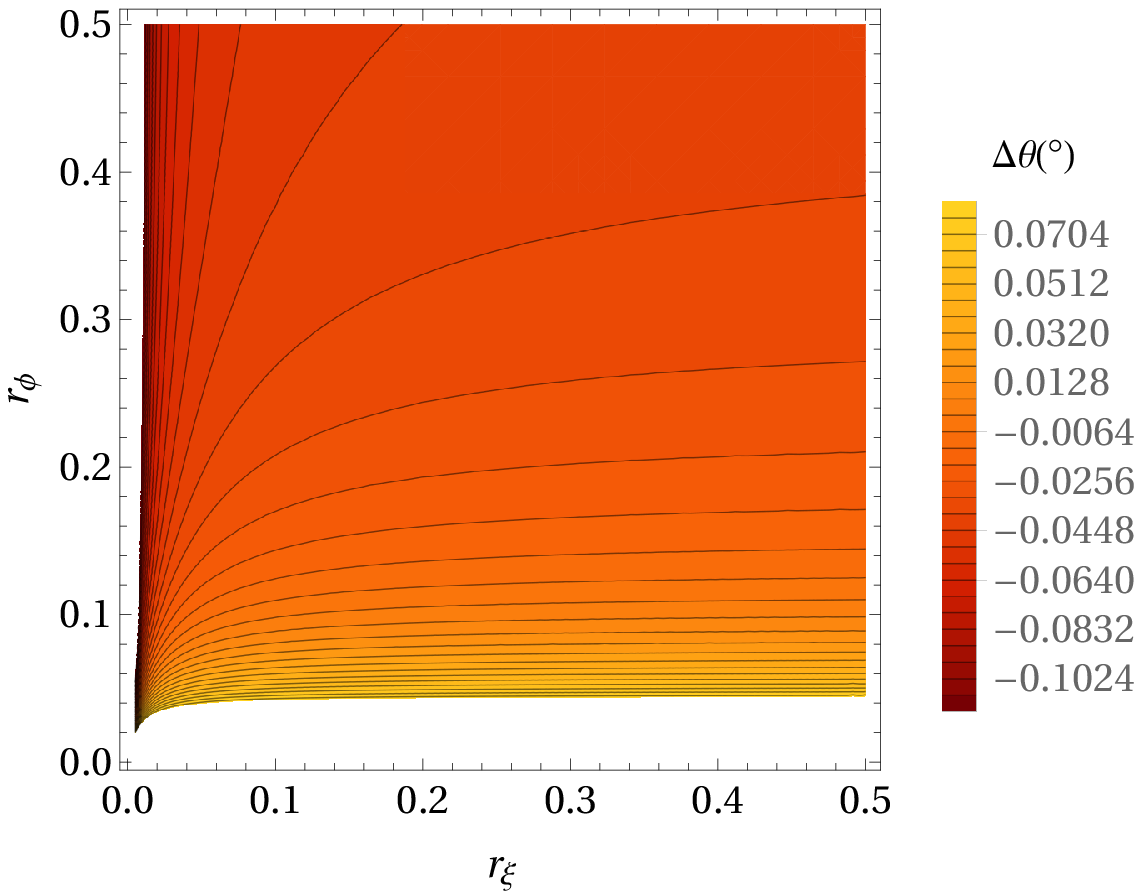}
\includegraphics[width=0.40\textwidth]{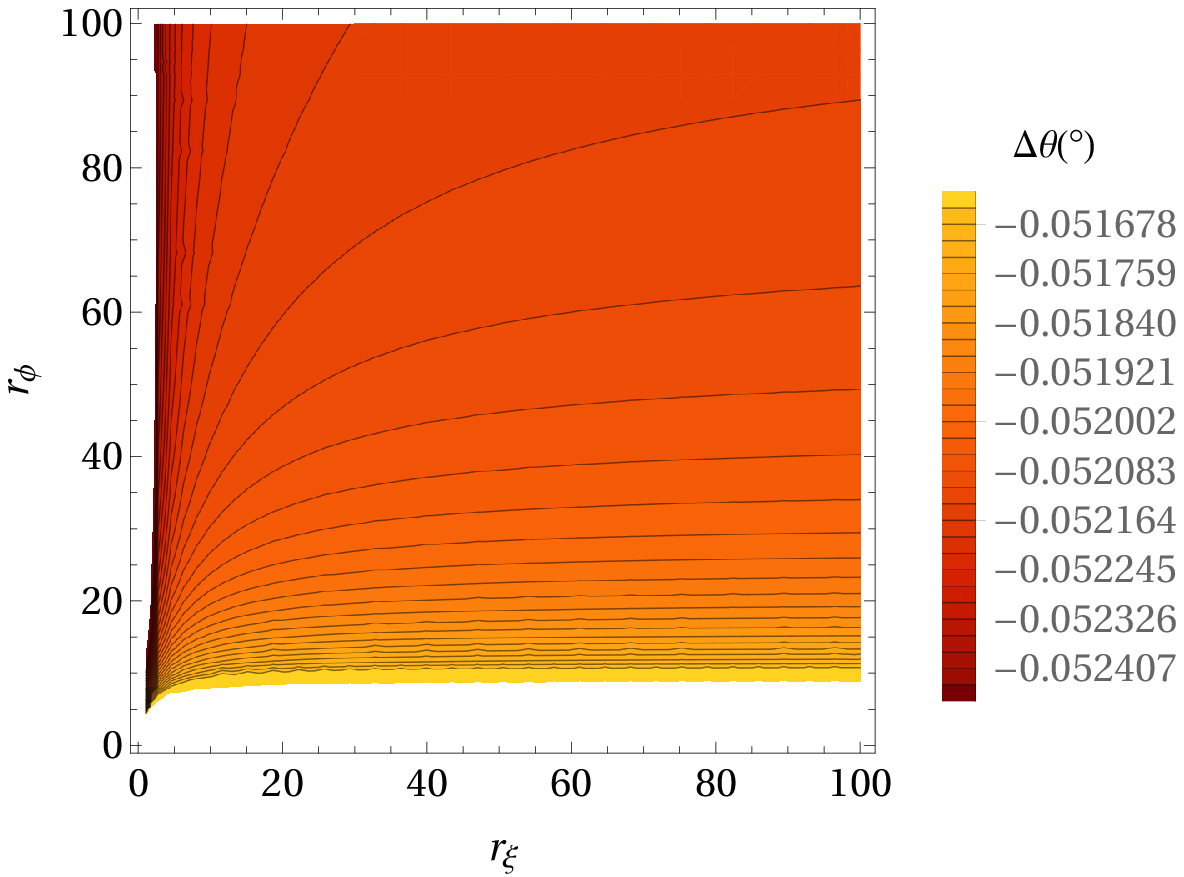}
\caption{(Color online) The precession per orbital period for
$r_{\phi}-r_{\xi}$ parameter space in the case of non-local modified 
gravity potential. With a decreasing value of angle of precession 
colors are darker.}
\label{fig03}
\end{figure*}

\begin{figure}[ht!]
\centering
\includegraphics[width=0.49\textwidth]{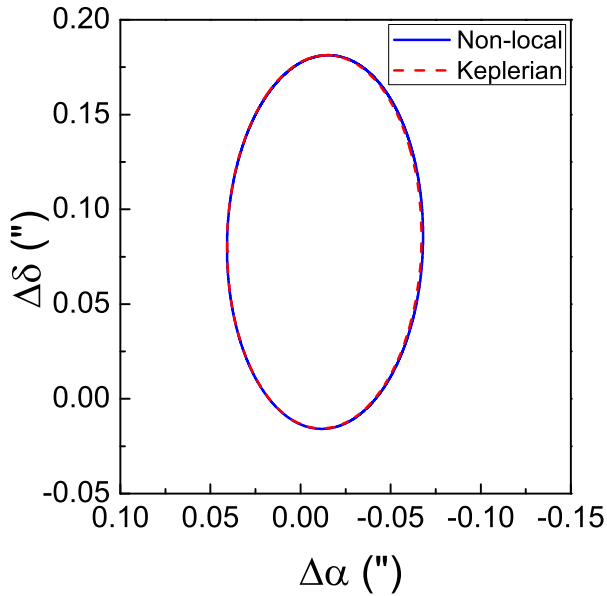}
\caption{Comparisons between the orbit of S2 star in Newtonian
gravity (red dashed line) and non-local gravity (blue solid line) in 
the observed plane, i.e. apparent orbit. Parameters of non-local gravity are $r_{\phi}$ = 1.2 AU and $r_{\xi}$ = 1.1 AU.}
\label{fig04}
\end{figure}

\begin{figure}[ht!]
\centering
\includegraphics[width=0.49\textwidth]{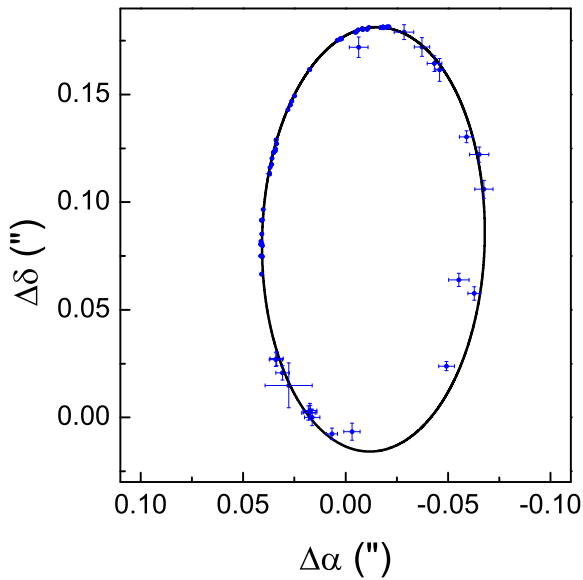}
\caption{A fitted orbit in non-local gravity through the 
following (parameters $r_{\phi}$ = 1.2 AU and $r_{\xi}$ = 1.1 AU (($\chi^{2}$ = 1.72)) observations of S2 star (denoted by points with error bars) NTT/VLT (see figure 3 from \cite{gill09a}).}
\label{fig05}
\end{figure}

\begin{figure*}[ht!]
\centering
\includegraphics[width=0.49\textwidth]{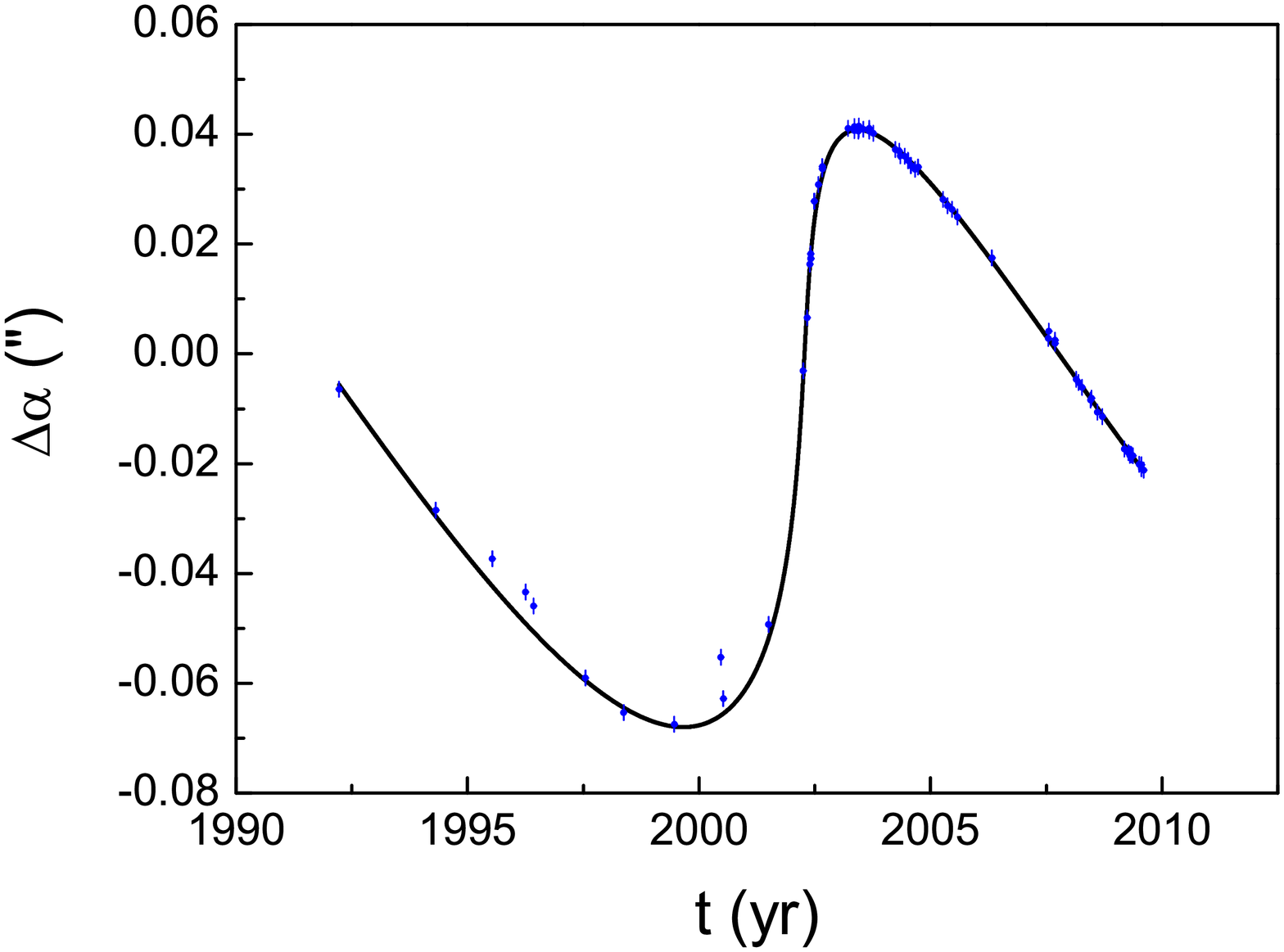}
\includegraphics[width=0.49\textwidth]{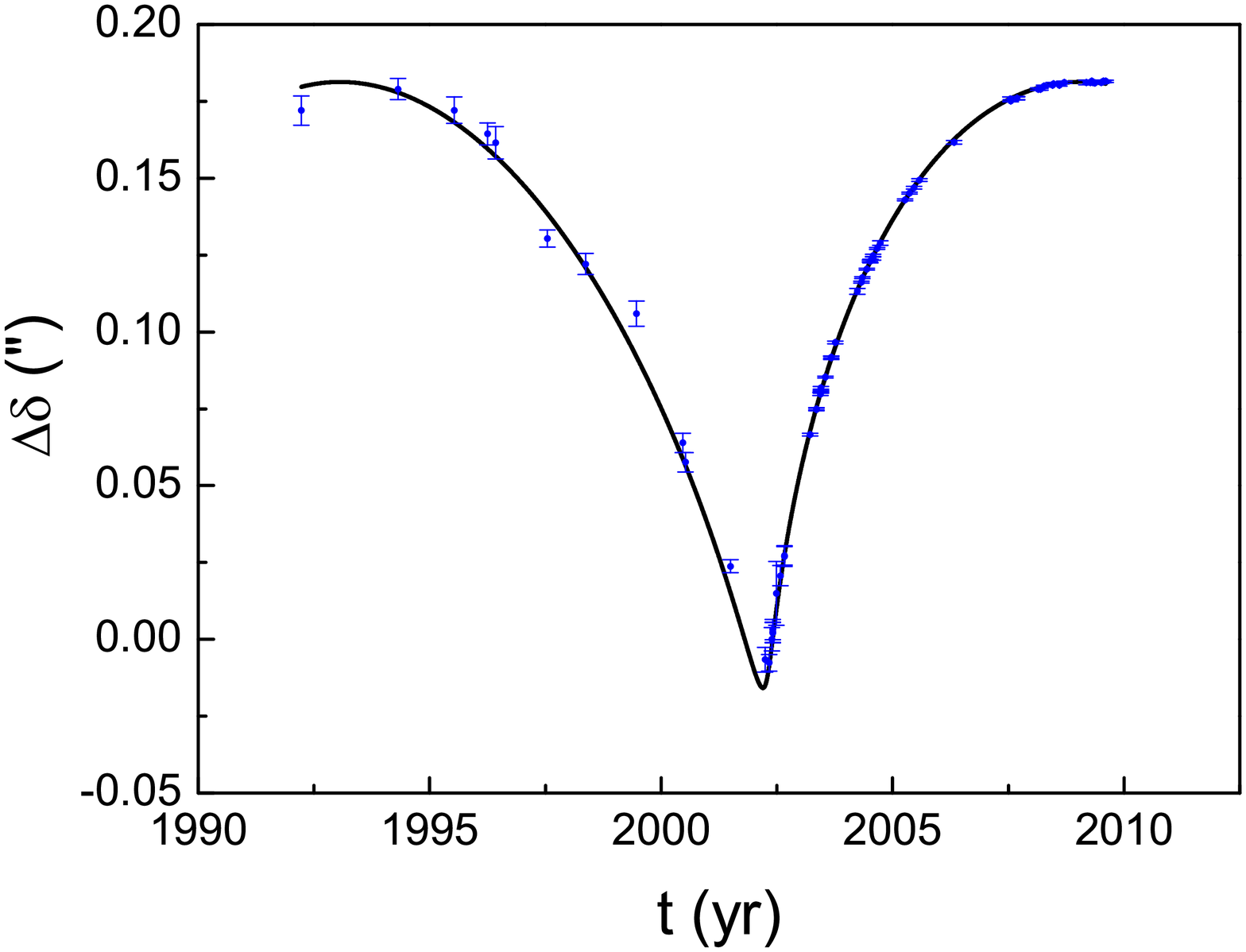}
\caption{The comparisons between the observed (circles with
error bars) and fitted (solid lines) coordinates of S2 star for 
$\Delta\alpha$ (left panel) and $\Delta\delta$ (right panel) in the case of NTT/VLT 
observations and non-local gravity potential (parameters $r_{\phi}$ = 1.2 AU and $r_{\xi}$ = 1.1 AU).}
\label{fig06}
\end{figure*}

In Figs. \ref{fig01} - \ref{fig02} we presented the maps of the 
reduced $\chi^{2}$ over the $r_{\phi}-r_{\xi}$ parameter space for 
all simulated orbits of S2 star which give at least the same or 
better fits than the Keplerian orbits. The second term of the RHS in Eq. (27) has a inverse $r$ term, namely $r^{-2} \times r = r^{-1}$. This term can potentially make a large deviation from the Keplerian orbit. A point is that the coefficient of this term is proportional to $18 r_{\xi} - 11r_{\phi}$. Therefore, the (probably) dominant deviation vanishes (and the $\chi ^2$ is thus small), if $r_{\phi} = (18/11) r_{\xi}$. This is exactly corresponding to the dark region (small $\chi^2$) in Figure 1. For more extended parameter space (see Figure 2), values of $\chi^2$ is almost nonsensitive on $r_{\xi}$ parameter.

As it can be seen from Fig. \ref{fig01} - Fig. \ref{fig02}, the most probable value for the 
scale parameter $r_{\phi}$, in the case of NTT/VLT 
data set observations of S2 star, is $\approx$ 0.1 - 2.5 AU. 
Moreover, as we see it is not possible to obtain constraints for the second 
length scale, $r_{\xi}$. This is because this length scale is 
associated with one of the scalar fields which is not dynamical, but 
it only plays an auxiliary role to localize the original non-local 
Lagrangian. Thus, it is obvious that we cannot constrain it. 

In order to calculate the orbital precession in non-local gravity, we
assume that the weak field potential does not differ significantly 
from the Newtonian potential, i.e. the perturbing potential:
\begin{equation}
V(r)=U_{NL}-U_N; \quad \text{ is small, where} \quad U_N=-\dfrac{GM}{r}.
\end{equation}
The weak field potential of the non-local gravity reads
\begin{align}
U_{NL} = &- \frac{G_N M}{r}\phi_c\ + 
\frac{G_N^2 M^2}{2c^2r^2} \left[\frac{14 }{9}\phi _c^2 
+ \frac{18 r_{\xi }-11 r_{\phi }}{6 r_{\xi } r_{\phi }} r \right]\nonumber \\
&+ \frac{G_N ^3 M^3}{2c^4r^3} \left[\frac{7 r_{\phi }-50 
r_{\xi }}{12 r_{\xi } r_{\phi }} \phi _c r-\frac{16 \phi 
_c^3}{27}+ \frac{2 r_{\xi }^2-r_{\phi }^2}{r_{\xi }^2 r_{\phi 
}^2}r^2\right]\,.
\end{align}

In Fig. \ref{fig03} we presented precession per 
orbital period for $r_{\phi}-r_{\xi}$ parameter space in the case of 
non-local gravity potential. We can notice that for values $r_{\phi}$ 
less then about $\approx$ 0.2 AU precession is positive, and for bigger 
values is negative. We hope that future more precise astronomical data 
will help us to better constrain non-local gravity parameters.

The particular form of the chosen Lagrangian among the class of 
non-local theories of gravity induces the precession of S2 star 
orbit. Depending of the values of parameters in the 
$r_{\phi}-r_{\xi}$ parameter space, precession of S2 star orbit 
calculated in non-local gravity can have positive or negative sign, 
i.e. the same or the opposite direction with respect to GR. In both 
cases the pericenter shift per orbital revolution is on the same order 
of magnitude as in GR, which predicts that pericenter of S2 star 
should advance by $0^\circ.18$ per orbital revolution \cite{gill09b}.

In Figs. \ref{fig04}-\ref{fig06}, we use one of the values for best fit parameters: $r_{\phi}$ = 1.2 AU and $r_{\xi}$ = 1.1 AU. For this choice of best fit papameters the value $\chi^{2}$ = 1.72. From Figs. \ref{fig01}-\ref{fig02} it is obvious that there are infinity number of such parameters where agreement is better than in Keplerian case ($\chi^{2}$ = 1.89), i. e. it is not possible to obtain reliable constrains on the parameter $r_{\xi}$. From Fig. \ref{fig03} (left panel) we can see that there are areas in the $r_{\phi}-r_{\xi}$ parameter space where precession of S2 star orbit 
calculated in non-local gravity can have positive or negative sign. In both cases of precession there are areas where agreement between non-local gravity and observation is better than in Keplerian case.  It means that one can make even stringer constrains of parameters $r_{\phi}$ and $r_{\xi}$ by requiring that precession must has positive or negative direction (like in GR or oposite). However, current precision of astrometric observations is not precise enough to definitly resolve this issue, and thus we give our result without this constraint. We choose area in the $r_{\phi}-r_{\xi}$ parameter space where precession is negative (opposite of GR) because in that case agreement with observations is better($\chi^{2}$ = 1.72) than in case when precession is positive ($\chi^{2}$ = 1.78). 

Comparison between the fitted orbit of S2 star in Newtonian gravity 
(red dashed line) and non-local gravity (blue solid line) in the 
observed plane is presented in Fig. \ref{fig04}. We can notice that 
difference between the orbit of S2 star in Keplerian case and in
non-local gravity is very small.

In Fig. \ref{fig05} the fitted orbit in non-local gravity through 
the NTT/VLT observations of S2 star (denoted by points with error 
bars) are presented. The comparisons between the observed 
(circles with error bars) and fitted (solid lines) $\Delta\alpha$ and 
$\Delta\delta$ coordinates of S2 star in the case of NTT/VLT observations and 
non-local gravity potential are given in Fig. \ref{fig06}. We can see that 
agreement between observed and fitted coordinates of S2 
star is very good.

\section{Conclusions}
\label{sec:conclusions}

Non-local gravity theories are very well motivated from cosmology, since they give a good explanation in the late-time acceleration of the Universe, without invoking exotic forms of matter-energy. However, a theory of gravity should be valid at all scales and that is why we wanted to study such theories at smaller scales, i.e. astrophysical. 

We considered a theory (1) proposed some years ago by Deser and Woodard [6], we “localized” it (3) as was proposed in [20] and we studied its invariance under point-transformations in a spherically symmetric spacetime. Surprisingly, we found that the forms of the distortion function that leave the action invariant are the same with those in a cosmological minisuperspace [31].

What we did next is, we selected the non-trivial form for the distortion function, i.e. the exponential $f(\phi) = 1+e^{\phi}$, that reproduces also the correct cosmological dynamics and we studied its weak field limit. After verifying that an asymptotically flat background consists a solution to the theory (3) with constant scalar fields, we perturbed the Minkowski background to $1/c^2$ terms up to third order, i.e. (19a)-(19d). The solutions we found are the Eqs. (24a)-(24d) and as we see two new-length scales arose; one for each scalar field.

We would like to confront our results with reality and specifically to find constraints on the two new length scales. That is why, we compared our results with the orbits of S2 star around the Galactic Center. We obtained the values for $r_{\phi}$ and $r_{\xi}$ parameters showing that the S2 star orbit in non-local gravity fits better the astrometric data than Keplerian orbit. The most probable value for the scale parameter $r_{\phi}$ is approximately from 0.1 to 2.5 AU. It is not possible to obtain reliable constrains on the parameter $r_{\xi}$ of non-local gravity using only observed astrometric data for S2 star because this length scale is associated with one of the scalar fields which is not dynamical, but only plays an auxiliary role to localize the original non-local Lagrangian.  

The precession of S2 star orbit in non-local gravity can 
have the same or the opposite direction with respect to GR, 
depending on the $r_{\phi}-r_{\xi}$ parameters, i.e. for values $r_{\phi} \lesssim 0.2$ AU, the precession is positive, and for bigger values is negative. The obtained orbital precession of the S2 star in non-local gravity is on the same order of magnitude as in GR; in the future, more precise astronomical data will help us better constrain the non-local gravity parameters. However, it is normal to believe that, non-local effects do not play a significant role at scales comparable to the S2 star orbit, i.e. astrophysical scales, but only at cosmological ones. There could be a screening effect, or a specific radius (maybe even given by the new length scale), after which non-local effects would start becoming significant. 

The approach we are proposing can be used to constrain  different
modified gravity models from stellar orbits around Galactic centre (see also \cite{Ivan1,Ivan2,Ivan3,Ivan4,dela18}).


\begin{acknowledgments}
The authors acknowledge the support of the Bilateral Cooperation 
between Serbia and Italy 451-03-01231/2015-09/1 ''Testing Extended 
Theories of Gravity at different astrophysical scales''. In addition, 
this work is partially supported by the COST Action CA15117 
(CANTATA) and ERASMUS+ Programme (for higher education student and staff mobility) between 
Dipartimento di Fisica, Universit\`{a} degli studi di Napoli 
''Federico II'' and Vin\v{c}a Institute of Nuclear Sciences, 
University of Belgrade. D.B., V.B.J. and P.J. acknowledge the support 
by Ministry of Education, Science and Technological Development of 
the Republic of Serbia, through the project 176003 ''Gravitation and 
the Large Scale Structure of the Universe''. KFD and SC acknowledge 
the support by the INFN sezione di Napoli (Iniziative Specifiche TEONGRAV and QGSKY). 
\end{acknowledgments}

\end{document}